\documentclass[twocolumn]{aastex701}

\defcitealias{Perdelwitz2024}{P24}

\begin{document}

\title{A Catalog of Homogeneously Derived Stellar Parameters for Spectroscopic Survey Stars}

\shorttitle{Homogeneous Stellar Parameters for Survey Stars}
\shortauthors{Stefanov et al.}

\correspondingauthor{Stefan Y. Stefanov}

\author[0000-0002-4993-2840]{Stefan Y. Stefanov}
\affiliation{Department of Astronomy, Sofia University "St Kliment Ohridski", 5 James Bourchier Blvd, BG-1164 Sofia, Bulgaria}
\affiliation{Institute of Astronomy and National Astronomical Observatory, Bulgarian Academy of Sciences, 72 Tsarigradsko shosse Blvd., 1784 Sofia, Bulgaria }
\email[show]{sstefanov@nao-rozhen.org}

\author[0000-0001-7663-4489]{Evelina Zaharieva}
\affiliation{Department of Astronomy, Sofia University "St Kliment Ohridski", 5 James Bourchier Blvd, BG-1164 Sofia, Bulgaria}
\email{ezaharieva@phys.uni-sofia.bg}

\author[0009-0005-9163-0095]{Djemma Ruseva}
\affiliation{Department of Astronomy, Sofia University "St Kliment Ohridski", 5 James Bourchier Blvd, BG-1164 Sofia, Bulgaria}
\email{ruseva@phys.uni-sofia.bg}

\author[0000-0002-5702-5095]{Milen Minev}
\affiliation{Institute of Astronomy and National Astronomical Observatory, Bulgarian Academy of Sciences, 72 Tsarigradsko shosse Blvd., 1784 Sofia, Bulgaria }
\email{msminev@nao-rozhen.org}

\author[0000-0001-6277-9644]{Denitza Stoeva}
\affiliation{Department of Astronomy, Sofia University "St Kliment Ohridski", 5 James Bourchier Blvd, BG-1164 Sofia, Bulgaria}
\email{dstoeva@phys.uni-sofia.bg}

\author[0000-0002-6859-0882]{Volker Perdelwitz}
\affiliation{Department of Earth and Planetary Science, Weizmann Institute of Science, Rehovot, Israel}
\email{volker.perdelwitz@weizmann.ac.il}

\author[0000-0003-3757-1440]{Lev Tal-Or}
\affiliation{Department of Physics, Ariel University, Ariel 40700, Israel}
\affiliation{ Astrophysics, Geophysics and Space Science Research Center, Ariel University, Ariel 40700, Israel}
\email{levtalor@ariel.ac.il}

\author[0000-0002-4947-144X]{Jerusalem T. Teklu}
\affiliation{Department of Physics, Ariel University, Ariel 40700, Israel}
\email{jerusalemt@msmail.ariel.ac.il}

\author[0000-0002-4872-7378]{Elena Vchkova Bebekovska}
\affiliation{Ss. Cyril and Methodius University in Skopje, Faculty of Natural Sciences and Mathematics-Skopje, Institute of Physics, Arhimedova 3 1000 Skopje, North Macedonia}
\email{elenavckova@gmail.com}

\author[0000-0003-1507-7230]{Desislava Antonova}
\affiliation{Department of Astronomy, Sofia University "St Kliment Ohridski", 5 James Bourchier Blvd, BG-1164 Sofia, Bulgaria}
\email{dantonova@phys.uni-sofia.bg}

\author[0000-0002-3117-7197]{Vladimir Bozhilov}
\affiliation{Department of Astronomy, Sofia University "St Kliment Ohridski", 5 James Bourchier Blvd, BG-1164 Sofia, Bulgaria}
\email{vbozhilov@phys.uni-sofia.bg}

\author[0000-0002-0236-775X]{Trifon Trifonov}
\affiliation{Department of Astronomy, Sofia University "St Kliment Ohridski", 5 James Bourchier Blvd, BG-1164 Sofia, Bulgaria}
\affiliation{Landessternwarte, Zentrum f\"ur Astronomie der Universt\"at Heidelberg, K\"onigstuhl 12, 69117 Heidelberg, Germany}
\email{trifonov@phys.uni-sofia.bg}

\begin{abstract}

Uniformly derived stellar parameters are vital for exoplanet demographic studies because they directly 
influence the inferred planetary masses, radii, and bulk densities. This study presents a new homogeneous 
catalog of physical stellar parameters for 5533 single stars observed by the HARPS, HIRES, and CARMENES 
radial-velocity (RV) surveys. Stellar parameters are determined using a Bayesian framework with two independent sets of stellar evolutionary models: MIST and PARSEC, 
using published spectroscopic 
parameter estimates and catalog values as priors. 
The resulting stellar effective temperatures, masses, radii, and surface gravities are compared and evaluated for consistency with stellar parameters listed in major exoplanet catalogs. While our estimates show consistency with those listed in external 
catalogs, we identify method-dependent differences in stellar masses for low-mass and 
pre-main-sequence stars, as well as G and K giants, where isochrone-based solutions may be affected by age--mass degeneracies. For low-mass stars such as M-dwarfs, 
the catalog provides masses derived from established mass--luminosity empirical relations, which tend to be more 
reliable. 

This catalog provides the largest uniformly derived stellar reference sample for Doppler survey targets and also illustrates the applicability and limitations of stellar-parameter homogenisation methods for future RV exoplanet demographics studies.

\end{abstract}


\section{Introduction} \label{sec:Intro}

Exoplanet detection techniques, such as the Doppler radial-velocity (RV) and transit methods, have led to the discovery of 6000 exoplanets as of 2026 June.
The characterization of these exoplanets is indirect, and it depends critically on the atmospheric and fundamental physical properties of their host stars. Stellar masses are required to convert RV semi-amplitudes into planetary minimum masses, while stellar radii are needed to translate transit depths into planetary radii \citep{Wright2013}. 
When both techniques are combined, the orbital inclination can be constrained, allowing the true planetary mass to be determined. This makes it possible to derive the planet's bulk density and thus place constraints on its internal composition. Accurate stellar masses and radii are also essential for strongly interacting 
multiple-planet systems, where $N$-body models applied to RV and transit data can be used to decode the dynamical architecture of the system \citep[e.g.,][]{vonStauffenberg2024, Vitkova2025, Almenara2026}. 
More generally, precise stellar parameters are required to infer reliable physical and orbital planetary properties, which are
important for the reverse engineering of planetary genesis.

Dedicated exoplanet databases, such as the Extrasolar Planets Encyclopaedia\footnote{\url{https://exoplanet.eu}} 
and the NASA Exoplanet Archive\footnote{\url{https://exoplanetarchive.ipac.caltech.edu/}}, provide basic stellar parameters together with the orbital and physical parameters of exoplanets. 
However, these stellar estimates are usually adopted from the corresponding discovery or follow-up papers. Differences in input physics, stellar models, and fitting methodologies can therefore introduce significant systematics, complicating comparative and statistical studies.
Homogeneous analyses are therefore essential, particularly for samples assembled from long-term Doppler programs that combine data obtained with different instruments, reduction pipelines, and analysis techniques. The importance of homogeneous stellar analyses has also been demonstrated by large spectroscopic population surveys. In the final DR5 release of the Gaia-ESO survey, \citet{Worley2024} combined the results of multiple independent pipelines through a Bayesian framework to provide homogeneous parameters for more than 80,000 FGK stars. This work shows that homogenization procedures can substantially reduce systematic differences between independent analyses. A successful effort toward a homogeneous analysis of exoplanet host stars was carried out within the ``Stars With ExoplanETs Catalogue'' \citep[SWEET-Cat;][]{Santos2013,Sousa2021,Sousa2024}, using the \textsc{ARES+MOOG} code \citep[see][and references therein]{Santos2013}. As of 2026, SWEET-Cat lists about 4400 unique exoplanet host stars, of which approximately 30\% have homogeneously derived stellar parameters, while the remainder rely on literature estimates. 

In this work, we present a stellar parameters catalog for spectroscopically observed stars monitored by state-of-the-art RV precision spectrographs.   
Adopting a homogeneous analysis, we derived stellar atmospheric and fundamental physical parameters for 5533 single stars observed by the HARPS \citep{Mayor2003}, HIRES \citep{Vogt1994}, and CARMENES \citep{Quirrenbach2014} spectrographs, ensuring internal consistency across the sample. The main aid of this catalog is the homogeneous orbital determination of exoplanet masses,
radii and bulk density, where possible, among other orbital parameters. We provide stellar parameters determined using a Bayesian framework with two independent sets of stellar evolutionary models: 
Mesa Isochrones and Stellar Tracks \citep[MIST,][]{Dotter2016,Choi2016} and PAdova and TRieste Stellar Evolution Code \citep[PARSEC,][]{Bressan2012}. For our analysis, we rely on previously published spectroscopic parameter estimates, primarily from our previous work in  \citet{Perdelwitz2024}, as priors rather than re-analyzing the spectra directly. When such estimates are not available, we adopt values from other catalogs with available stellar-parameter measurements. We cross-check our estimates with those listed in the external catalogs and empirical relations, using these comparisons to assess the range of validity, applicability, and limitations of the two independent stellar-modeling methods adopted in this work. Our catalog provides the largest uniformly derived stellar reference sample for RV and transit planet hosts, serving as a robust analysis for future exoplanet demographics studies.

The paper is organized as follows: In \autoref{sec:selection} and \autoref{sec:Methods}, we describe the sample selection and analysis methods. In \autoref{sec:Results}, we compare our results with existing catalogs to assess the consistency of our measurements. Finally, in \autoref{sec:highlight} we summarize the main results and conclusions.

\section{Selection of Targets}\label{sec:selection}

The core sample of 4200 stars in this work is drawn from the {\sc HARPS-RVB}ank compiled by \citet{Trifonov2020} and extended by \citet{Perdelwitz2024}. This database provides homogeneous, publicly available RV and stellar activity index measurements derived from archival spectra obtained with the High Accuracy Radial velocity Planet Searcher \citep[HARPS;][]{Mayor2003}, mounted on the 3.6\,m telescope at La Silla Observatory and operated by the European Southern Observatory. It is a fiber-fed, cross-dispersed echelle spectrograph with a high resolving power ($R \approx 115,000$), long-term RV stability, and sub $\mathrm{m}\,\mathrm{s^{-1}}$ RV precision. The {\sc HARPS-RVB}ank contains spectra that have been re-processed using the \texttt{SERVAL}\footnote{\url{https://github.com/mzechmeister/serval}} \citep{Zechmeister2018} pipeline, which derives RVs through template matching and achieves improved precision relative to the standard {\sc HARPS} data reduction system. In addition to RVs, \texttt{SERVAL} provides activity indicators such as H$\alpha$, Ca II H\&K, and chromatic indices. The {\sc HARPS-RVB}ank database aims to provide a uniform data product that can be readily used for independent analyses without requiring access to the raw spectra. 

A total of 1616 targets are included in the HIRES/Keck RV catalog compiled by \citet{Butler2017}. The High Resolution Echelle Spectrometer \citep[HIRES;][]{Vogt1994} is mounted on the 10\,m Keck~I telescope at the W.~M.~Keck Observatory and is optimized for high-precision spectroscopic studies. HIRES has achieved long-term RV precision at the few $\mathrm{m}\,\mathrm{s^{-1}}$ level using the iodine absorption cell technique for wavelength calibration \citep[e.g.,][]{Butler1996}. More recently, \citet{Teklu2025} compiled precision RV measurements derived from the full set of publicly available HIRES spectra obtained with the iodine cell, spanning more than two decades of observations. The updated catalog contains 78,920 RV measurements and includes Ca\,{\sc ii} H\&K chromospheric activity indicators.

Additionally, we include stars from the Calar Alto high-Resolution search for M dwarfs with Exo-Earths with Near-infrared and optical Échelle Spectrographs \citep[CARMENES,][]{Quirrenbach2014} guaranteed-time survey
\citep{Reiners2018,Ribas2023}. The {\sc CARMENES} instrument is installed at the 3.5\,m telescope of the Calar Alto Observatory and is operated by a German–Spanish consortium. It simultaneously covers the visible ($R \approx 94,600$) and near-infrared ($R \approx 80,400$) wavelength ranges and is designed to achieve $\mathrm{m}\,\mathrm{s^{-1}}$ RV precision for the detection of low-mass planets around low-mass stars. The 357 stars in this subsample are nearby M dwarfs spanning spectral types M0.0\,V to M9.5\,V, selected to optimize RV sensitivity \citep{Ribas2023}.

Some of the targets in this combined sample are binaries, while all of our analysis is tailored for single stars. To avoid binary contamination, we queried the SIMBAD Astronomical Database\footnote{\url{https://simbad.cds.unistra.fr/simbad/}} and flagged 525 stars with object types assigned as SB* or EB*. These two types represent spectroscopic and eclipsing binaries, and we exclude them from all following analyses.   

In total, our catalog contains 5533 single stars, of which 524 are present in more than one of the RV databases considered here.

\section{Methodology}\label{sec:Methods}

We derive fundamental and atmospheric parameters of stars using tools based on two different stellar model libraries: the MESA Isochrones and Stellar Tracks \citep[MIST;][]{Dotter2016, Choi2016} and the Padova and Trieste Stellar Evolution Code used to compute stellar evolutionary tracks \citep[PARSEC;][]{Bressan2012}. This was done using two independent software packages: The first, \texttt{astroARIADNE}, is a publicly available\footnote{\url{https://github.com/jvines/astroARIADNE}}
 Python implementation of the Bayesian spectral energy distribution (SED) fitting framework introduced by \citet[][{\sc ARIADNE}: spectrAl eneRgy dIstribution bAyesian moDel averagiNg fittEr]{Vines&Jenkins2022}. This package fits SEDs using the following input parameters: effective temperature ($T_{\mathrm{eff}}$), surface gravity ($\log g$), metallicity [M/H], extinction ($A_V$), stellar radius ($R_\star$), and distance. The tool then performs isochrone fitting using MIST models to estimate stellar masses and ages. The second code we use, \texttt{SPOG+} \citep[Stellar Parameters of Giants and more;][]{Stock2018}, is a publicly available\footnote{\url{https://github.com/StephanStock/SPOG}}
 package that derives stellar parameters using PARSEC evolutionary models. The method requires four input parameters: magnitudes in two photometric bands, initial metallicity guess, and stellar parallax. In this section, we describe our implementation of \texttt{ARIADNE} and \texttt{SPOG+}, as well as the prior parameters adopted as inputs for these tools. 

\subsection{Prior Stellar Parameters}\label{sec:prior_sp}

We cross-matched all targets in our sample with the third data release of Gaia \citep[Gaia DR3;][]{DR3_2023}. For each star, we provide the DR3 and DR2 source identifiers, parallax, proper motion, $G$-band magnitude, $BP-RP$ color, and the Gaia variability flag (\texttt{VarFlag}). We also include the stellar parameters available in DR3 derived with the GSP-Aeneas algorithm \citep{Andrae2023}. This module estimates stellar parameters for single stars using the low-resolution $BP/RP$ spectra together with Gaia astrometry and $G, BP$, and $RP$ photometry. For stars with $G \lesssim 19$, it provides estimates of $T_{\mathrm{eff}}$, $\log g$, [M/H], extinction, absolute magnitude, radius, and distance using an ensemble Markov Chain Monte Carlo sampler. All of these data products are available through the official Gaia data release\footnote{\url{https://www.cosmos.esa.int/web/gaia/data-release-3}}.

Another source of stellar parameters is the latest version of the TESS Input Catalogue \citep[TICv8.2;][]{tessInputCatalogue}. This all-sky catalog was developed to support target selection and characterization for the Transiting Exoplanet Survey Satellite \citep[TESS;][]{Ricker2015}. It combines astrometric and photometric measurements from numerous space missions and ground-based surveys to provide homogeneous estimates of stellar properties, including effective temperature, radius, mass, luminosity, and surface gravity. Different methodologies are applied to distinct stellar populations (e.g., dwarfs, subgiants, and giants) to mitigate systematic biases. We cross-matched our sample with TICv8.2 using the Gaia DR2 identifiers of each star and obtained catalog values for 5372 single stars.

For 3156 of our targets observed with {\sc HARPS}, stellar parameters were derived by \citet{Perdelwitz2024}, hereafter \citetalias{Perdelwitz2024}. This was achieved using the code \texttt{SPECIES}\footnote{\url{https://github.com/msotov/SPECIES}}
 \citep[Spectroscopic Parameters and atmosphEric ChemIstriEs of Stars;][]{speciesI_2018, speciesII_2021}. This tool measures fundamental stellar parameters directly from spectra, using equivalent widths of absorption lines. Input templates are constructed by co-adding spectra to achieve high signal-to-noise ratios. For stars with template \mbox{S/N $> 20$} \texttt{SPECIES} provides estimates of $T_{\mathrm{eff}}$, $\log g$, [M/H], and projected rotational velocity ($v\sin i$).

When compared to TICv8.2, the estimates of $T_{\mathrm{eff}}$ and $\log g$ from \texttt{SPECIES} contain several clusters of outliers (see Figures~3–5 of \citetalias{Perdelwitz2024}). We approach these discrepancies by introducing an additional quality flag for these stars. Most of the objects we flagged have unjustifiably assigned temperatures around $T_{\mathrm{eff}} \simeq 5500\,\mathrm{K}$ and $T_{\mathrm{eff}} \simeq 9000\,\mathrm{K}$ by \texttt{SPECIES}. We also identify stars clustered at $\log g \approx 4.90$ and $\log g \approx 4.36$ that are inconsistent with both TICv8.2 and Gaia GSP-Aeneas. In addition, \texttt{SPECIES} is known to converge to unreliable solutions for stars with $T_{\mathrm{eff}} \lesssim 4300\,\mathrm{K}$ \citep{speciesI_2018, speciesII_2021}, and these objects are also flagged. We do not use any of the \citetalias{Perdelwitz2024} parameters for flagged stars as input in our analysis later on.

\section{Results and Comparisons}\label{sec:Results}

\subsection{Stellar Parameters Using \texttt{astroARIADNE}}

\texttt{ARIADNE} fits multi-band photometric SEDs to a suite of stellar atmosphere models using a Bayesian model-averaging framework. This approach combines information from multiple model grids and helps mitigate systematic biases associated with individual models. 
The input photometry is automatically retrieved from multiple catalogs available through VizieR\footnote{\url{http://vizier.u-strasbg.fr/}}
 and MAST\footnote{\url{https://archive.stsci.edu/}}. The photometry retrieval procedure is described in detail in \citet{Vines&Jenkins2022}. For the SED analysis, we use the atmospheric model grids PHOENIX v2 \citep{Husser2013}, BT-Settl, BT-NextGen, and BT-Cond \citep{Hauschildt1999, Allard2012}, as well as the Castelli \& Kurucz \citep{Castelli&Kurucz2003} and Kurucz \citep{Kurucz1993} models.

We adopt the following priors for the fitting procedure: A normally distributed prior is assigned to the distance using the Gaia EDR3 estimate from \citet{Bailer-Jones2021}, with a standard deviation equal to five times the largest reported uncertainty. For interstellar extinction, we adopt a uniform prior ranging from zero to the maximum line-of-sight extinction, calculated from the Galactic dust map of \citet{Schlafly&Finkbeiner2011} using the Python package \texttt{dustmaps}\footnote{\url{https://github.com/gregreen/dustmaps}}
 \citep{Green2018}. For the stellar parameters $T_{\mathrm{eff}}$, $\log g$, radius, and [Fe/H], we construct normal priors based on the estimates provided by \citetalias{Perdelwitz2024} in the {\sc HARPS-RVB}ank for stars not associated with a poor-quality flag. For stars without such estimates, we instead adopt the corresponding values from TICv8.2, and for those lacking TIC entries, we use the Gaia DR3 GSP-Aeneas estimates. In each case, the standard deviation of the prior is set to five times the largest reported uncertainty of the corresponding parameter. For stars without previous estimates of these parameters, we adopt the default \texttt{ARIADNE} priors, except for the stellar radius, for which we extend the upper bound from $100\,R_\odot$ to  $500\,R_\odot$. This was done because a few of the evolved stars in our sample have reported radii 
of up to $ \sim 230\,R_\odot$ by TICv8.2. 

\begin{figure}[t]
\includegraphics[width=\columnwidth]{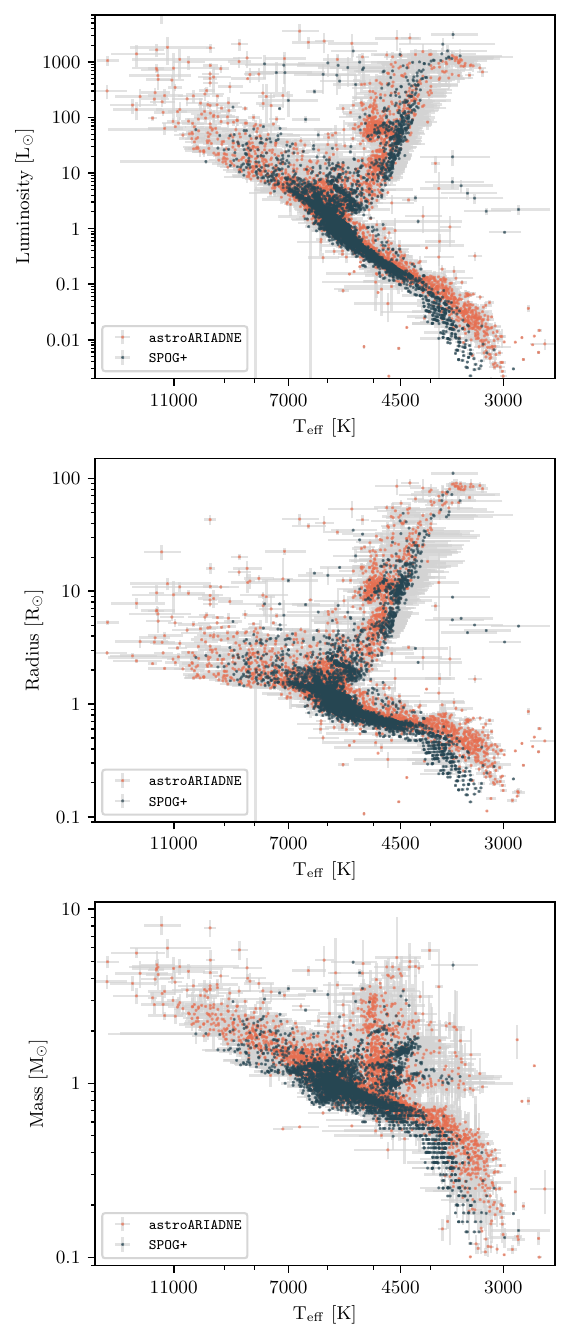}
\caption{Top panel: Hertzsprung--Russell diagram constructed using the computed effective temperatures and luminosities with \texttt{ARIADNE} (orange dots) and \texttt{SPOG+} (dark blue dots). Middle panel: temperature versus radius for all of the stars computed with \texttt{ARIADNE} and \texttt{SPOG+}. Bottom panel: temperature versus mass for all of the stars computed with \texttt{ARIADNE} and \texttt{SPOG+}.
\label{fig:g2hr}}
\end{figure}

 We successfully ran \texttt{ARIADNE} for 5014 single stars in our sample. The code outputs $T_{\mathrm{eff}}$, $R_\star$, $M_\star$, Age, $\log g$, [Fe/H], distance, $A_V$, $L_\star$, and EEP\footnote{Equivalent Evolutionary Point; see \autoref{sec:low_mass}.} for each target. A Hertzsprung--Russell diagram, $R_\star$ versus $T_{\mathrm{eff}}$, and $M_\star$ versus $T_{\mathrm{eff}}$ diagrams constructed from the retrieved parameters are shown with orange points in \autoref{fig:g2hr}.

\subsection{Stellar Parameters Using \texttt{SPOG+}}

We also used the code \texttt{SPOG+} \citep[Stellar Parameters of Giants and more;][]{Stock2018} to compute stellar parameters. The \texttt{SPOG+} framework was originally developed to improve parameter accuracy and evolutionary stage classification for a sample of 372 giants from the Lick planet search \citep{Frink2001, Reffert2015}. It can be used to derive fundamental stellar parameters for stars that have evolved off the main sequence and are on the red giant branch (RGB) or the core-helium-burning (horizontal branch, HB) evolutionary phases. It is optimized to provide mass estimates for low- and intermediate--mass stars (giants and subgiants), for which classical isochrone fitting often produces ambiguous solutions due to overlapping evolutionary tracks in the Hertzsprung–Russell diagram. The plus version of the code employs the same Bayesian inference framework and is also suitable for main-sequence and pre-main-sequence stars. \texttt{SPOG+} incorporates the evolutionary tracks of \citet{Bressan2012} and spans wide ranges in stellar masses ($0.09$–$12\,M_\odot$) and metallicity ([Fe/H] $\approx -1.5$ to $+0.7$).

The \texttt{SPOG+} tool requires four input parameters: magnitudes in two photometric bands, an initial metallicity estimate, and the stellar parallax. For the photometric bands, we adopt the 2MASS $J$ and $K_s$ magnitudes, while parallaxes are taken from \citet{Bailer-Jones2021}. The metallicity estimates are adopted from \citetalias{Perdelwitz2024} for stars with spectroscopic measurements; for the remaining stars, we use values from TICv8.2 or Gaia DR3 GSP-Aeneas. We do not model extinction or reddening in the \texttt{SPOG+} analysis, as the majority of stars in our sample lie within 250\,pc. 

As output, the tool provides $M_\star$, $R_\star$, $\log g$, Age, $L_\star$, $T_{\mathrm{eff}}$, and the probability of the target being in each of the evolutionary stages --- pre-main-sequence, main-sequence, RGB, and HB. The resulting subsample of 4018 single stars with parameters derived using this tool is shown in \autoref{fig:g2hr}, indicated by dark-blue points.

\subsection{Comparison With Other Databases}\label{sec:floats}

\begin{figure*}[p]
\includegraphics[width=\textwidth]{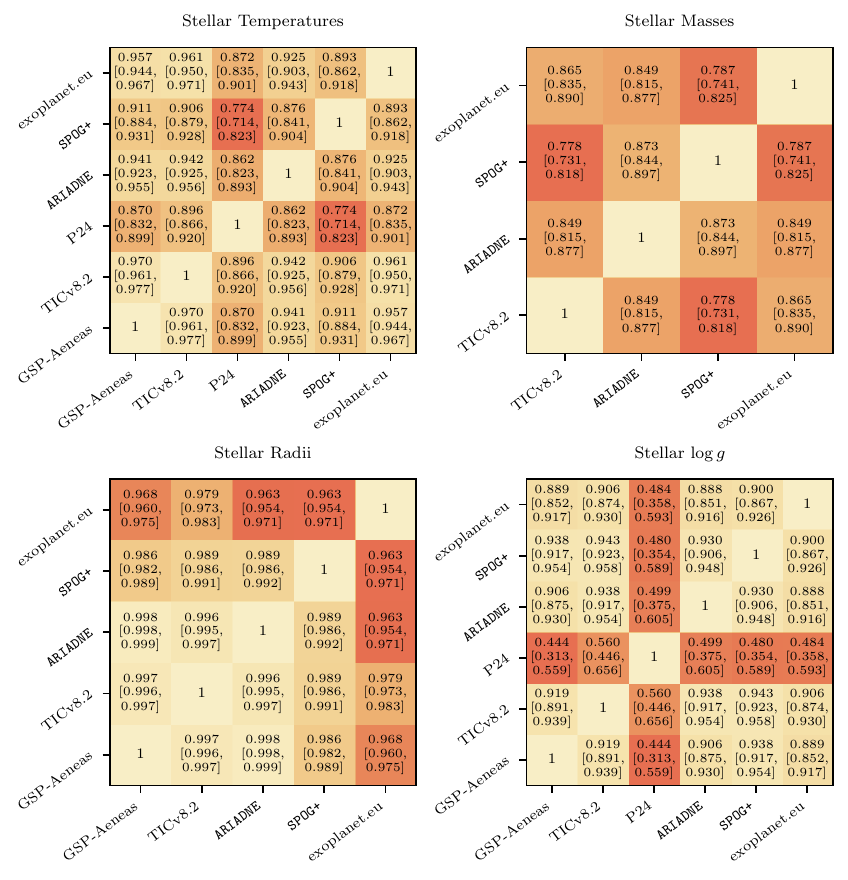}
\caption{Each panel shows the pairwise Pearson correlation coefficients computed using only stars with valid measurements in all catalogs considered in this work (Gaia DR3 GSP Aeneas, TICv8.2, \citetalias{Perdelwitz2024}, \texttt{ARIADNE}, \texttt{SPOG+}, and the \url{exoplanet.eu} database). The annotated values correspond to the correlation coefficient along with 95\% confidence intervals. The correlation matrices were constructed using 212 stars for $T_{\text{eff}}$, \mbox{320 for $M_\star$}, \mbox{278 for $R_\star$}, and \mbox{166 for $\log{g}$}. Color shading reflects the magnitude of the correlation coefficient and is scaled for each panel separately. 
\label{fig:correls}}
\end{figure*}

In addition to our estimates of \texttt{ARIADNE} and \texttt{SPOG+}, our catalog contains parameters from several external sources: the Gaia DR3 GSP-Aeneas pipeline, the TICv8.2 pipeline, and spectroscopic estimates from \citetalias{Perdelwitz2024}. We also compare our results with previously published stellar parameters, taken from the database available at \url{exoplanet.eu}\footnote{We used a version of the database dating from 2026 February.}. In this database, stellar parameters are compiled from published, peer-reviewed studies and are not recalculated by the database itself. For each star, the catalog provides effective temperature, surface gravity, metallicity, stellar mass, and radius derived from a variety of techniques, including spectroscopic analyses, transit and RV modeling, stellar evolution modeling, and astrometric constraints.

The comparison shows good agreement between the \url{exoplanet.eu} database and our estimates for solar-like stars. For late-type main-sequence stars, there are systematics present in effective temperature and mass, and for some giant stars, mass and radius are poorly constrained. In the following subsections, we attempt to characterize these method-dependent differences for several atmospheric fundamental parameters relevant for exoplanet characterization. For the stellar effective temperatures, masses, radii, and surface gravities, we performed a comparative analysis using all databases discussed in \autoref{sec:prior_sp}. Each parameter was treated independently, considering only stars with measurements available in every database: 212 stars for $T_{\mathrm{eff}}$, \mbox{320 for $M_\star$}, \mbox{278 for $R_\star$}, and \mbox{166 for $\log g$}. For each pair of databases, we computed the Pearson correlation coefficient \citep{Pearson1895} and derived the corresponding 95\% confidence intervals using the Fisher $z$-transformation. The resulting correlation matrices are shown in \autoref{fig:correls}.

Because this comparison requires target cross-match among all databases, the resulting samples are significantly smaller than the full stellar sample analyzed in this work. The sub-sample sizes are primarily driven by the coverage of the \url{exoplanet.eu} database, which inherit selection biases associated with known exoplanet host stars and the observational techniques used to detect them. Most surveys preferentially observe bright, nearby, and relatively inactive stars, leading to an over-representation of FGK-type stars and a bias against young, active, or intrinsically faint stellar populations. 
Within our correlation-analysis sample, cool main-sequence stars and evolved stars are therefore under-represented. To mitigate these limitations, we also perform a separate, more detailed comparison for each stellar parameter using the full set of overlapping stars for every pair of catalogs. To quantify systematic offsets and slopes in these comparisons between databases, we perform linear regressions using the orthogonal distance regression algorithm implemented in the Python package \texttt{scipy} \citep{scipy_2020}.

\subsection{Stellar Effective Temperatures}

The upper-left panel of \autoref{fig:correls} shows the pairwise correlation coefficients for the 212 single stars that have temperature estimates available in all databases used in this work. Lower correlations between the catalog of \citetalias{Perdelwitz2024} and the other databases remain, even after removing the flagged spectroscopic outliers. This effect is most pronounced in the comparison with \texttt{SPOG+}, for which the Pearson correlation coefficient is 0.774 --- the lowest among any catalog pair. Overall, the different catalogs show good agreement in effective temperature. However, it should be noted that the stars included in the correlation analysis are predominantly solar-type main-sequence stars and are therefore not fully representative of the broader stellar sample considered in this catalog.

\begin{figure}[t!]
\includegraphics[width=\columnwidth]{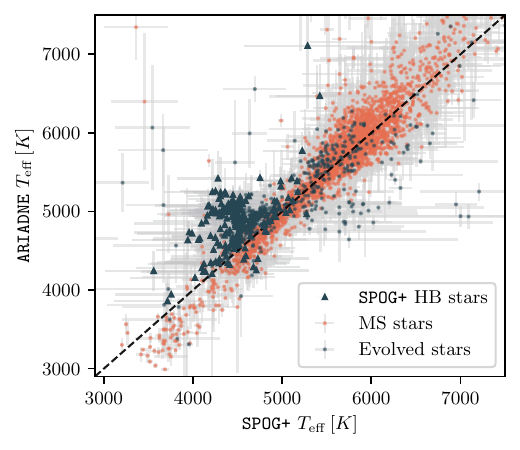}
\caption{Comparison of stellar effective temperatures derived with \texttt{ARIADNE} and \texttt{SPOG+}. Blue circles denote stars classified as evolved by both methods, while blue triangles indicate stars identified as horizontal branch objects by both. Orange circles represent stars that are consistently classified as main-sequence by both \texttt{ARIADNE} and \texttt{SPOG+}.
\label{fig:teff_comp}}
\end{figure}

Overall, we find good agreement in effective temperature when comparing the 4180 stars common to \texttt{SPOG+} and \texttt{ARIADNE}. For the bulk of main-sequence stars in our sample, we detect no significant slopes or systematic offsets. The comparison can be seen on \autoref{fig:teff_comp}. However, there are two clusters of stars present, deviating from the 1:1 black dashed line. The first group consists of cool main-sequence stars with $T_{\mathrm{eff}} \lesssim 4500$\,K, for which \texttt{ARIADNE} returns temperatures that are on average $330 \pm 55$\,K cooler than those obtained with \texttt{SPOG+}. The second group includes stars identified by \texttt{SPOG+} as evolved objects with probabilities exceeding 90\% being located on the horizontal branch. For these stars, \texttt{ARIADNE} yields temperatures that are $510 \pm 51$\,K higher on average than the \texttt{SPOG+} estimates, although both methods agree on their evolutionary classification. These offsets were determined by fixing the slopes of the linear regressions to unity. 

When compared with spectroscopically determined temperatures, both \texttt{SPOG+} and \texttt{ARIADNE} show no significant systematic offsets, except for the flagged stars lacking reliable spectroscopic parameters (see \autoref{sec:prior_sp} and Figures~3–5 of \citetalias{Perdelwitz2024}).

\subsection{Stellar Masses}\label{sec:low_mass}

\texttt{ARIADNE} provides two mass estimates for each target. The first is the gravitational mass calculated using the relation $\log{M_\star} = \log{g} + 2\log{R_\star} - 4.437$, where $\log g$ and $R_\star$ are derived from the SED fitting of stellar atmosphere models. Because SED fitting is only weakly sensitive to $\log g$, this estimate depends strongly on the adopted priors. The second method determines stellar masses by interpolating MIST isochrones within the Bayesian model-averaging framework of \texttt{ARIADNE}. As this approach does not rely on the $\log g$ prior, we adopt the isochrone-based masses for all subsequent analyses. 

We first compare the masses derived with \texttt{ARIADNE} and \texttt{SPOG+} to those reported in TICv8.2. There are 4931 single stars in common between TICv8.2 and \texttt{ARIADNE}, and 3967 single stars in common between TICv8.2 and \texttt{SPOG+}. In both cases, we identify systematic offsets that depend on the evolutionary stage of the stars. To quantify these effects, we divide the sample into three groups: pre-main-sequence (PMS), main-sequence (MS), and evolved stars.
For the classification based on \texttt{ARIADNE}, we use the Equivalent Evolutionary Point (EEP) parameter, which labels physically equivalent stages of stellar evolution across tracks of different initial masses. Rather than using stellar age directly --- which is highly mass dependent and nonlinear --- the MIST models re-parametrize evolutionary tracks in terms of EEPs to ensure well-behaved interpolation between tracks \citep{Dotter2016, Choi2016}. Using this parameter, we classify 544 stars with $\mathrm{EEP} \leq 202$ as PMS stars, 3546 stars with $202 < \mathrm{EEP} \leq 454$ as main-sequence stars, and 214 stars with $\mathrm{EEP} > 454$ as evolved stars.  

\texttt{SPOG+} provides evolutionary-stage probabilities for each star corresponding to the PMS, MS, RGB, and HB phases. For the comparison with TICv8.2 we select 2292 stars classified as MS with probability $>90\%$, 83 PMS stars with probability $>90\%$, and 163 evolved stars (RGB or HB) with probability $>90\%$. 

\begin{figure}[ht!]
\includegraphics[width=\columnwidth]{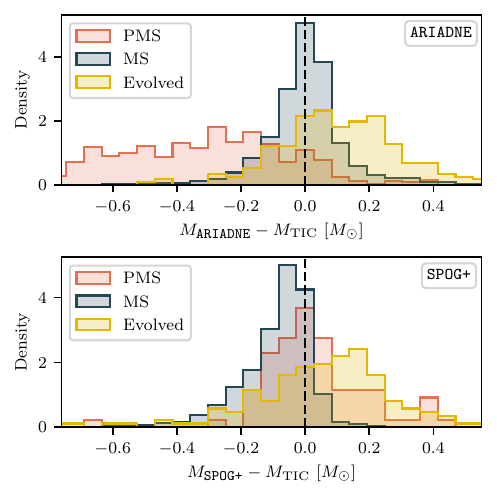}
\caption{Comparison between stellar masses from TICv8.2 ($M_{\mathrm{TIC}}$) and those calculated with \texttt{ARIADNE} and \texttt{SPOG+}. The two panels show density histograms of the residuals relative to TICv8.2 for stars at different evolutionary stages. The full sample of stars in common is divided into three groups: pre-main-sequence (PMS), main-sequence (MS), and evolved stars.
\label{fig:tic_mcomp}}
\end{figure}

The differences between TICv8.2 masses and those derived with \texttt{SPOG+} and \texttt{ARIADNE} are shown in  \autoref{fig:tic_mcomp}. 

For the masses by \texttt{ARIADNE}, there are significant offsets for both PMS and evolved stars. To investigate the PMS discrepancies, we cross-matched these stars with the \url{exoplanet.eu} database. We identified 19 stars with available mass estimates, of which 10 show significant disagreement. This likely reflects the age–mass degeneracy in isochrone fitting that can affect young stars, leading \texttt{ARIADNE} to assign underestimated masses to some low-mass targets. Evolved stars show a systematic offset of $\sim 0.15\,M_\odot$ in both \texttt{ARIADNE} and \texttt{SPOG+} relative to TICv8.2, with both methods yielding higher masses. We do not find this offset in comparisons with the \url{exoplanet.eu} database, which suggests this offset could be intrinsic to TICv8.2. 

Unlike \texttt{ARIADNE}, masses of PMS stars from \texttt{SPOG+} show good agreement with TICv8.2. For MS stars, however, the residual distribution is skewed, with \texttt{SPOG+} mass estimates being, on average, $0.09\, M_\odot$ higher than those reported in TICv8.2 for the same stars.

\subsubsection{Low-mass Stars}

The offsets relative to TICv8.2 are large enough to be significant for low-mass stars, particularly if they are incorrectly classified as PMS objects. To further investigate this mass regime, we computed masses for all stars with absolute magnitudes in the $K_s$ band from 4.5 to 10.5\,mag using the \texttt{M\_-M\_K-} code\footnote{\url{https://github.com/awmann/M_-M_K-}} by \citet{Mann2019}. This method estimates stellar masses for dwarf stars in the mass range $0.075$–$0.70\,M_\odot$ with 2MASS $K_s$ magnitudes and Gaia parallaxes as input. The underlying $M_{K_s}$–$M_\star$ relation is calibrated on a sample of 62 binary systems with orbits solved by direct imaging astrometric techniques. Using the total masses of these binaries, their component resolved $K_s$ magnitudes, and system parallaxes, \texttt{M\_-M\_K-} employs a Bayesian sampling framework to produce posterior mass distributions for single stars. The relation provides stellar masses with a typical accuracy of $\sim2\% - 3\%$ and is largely insensitive to stellar metallicity.

As an additional comparison, we also use the mass–luminosity relation of \citet{Trifonov2018}, calibrated using the mass estimates from \citet{Benedict2016} and \citet{Delfosse2000}. The relation uses 2MASS $K_s$ magnitudes as input and has the form $M = a_0 + a_1K_s + a_4K_s^4$. Here the coefficients are \mbox{$a_0$ = (1.735$\pm$0.005)}, \mbox{$a_1$ = $(-$2.262$\pm0.016) \times 10^{-1}$}, and \mbox{$a_4$ = (6.2$\pm1.6)\times10^{-5}$}.

\begin{figure}[t!]
\includegraphics[width=\columnwidth]{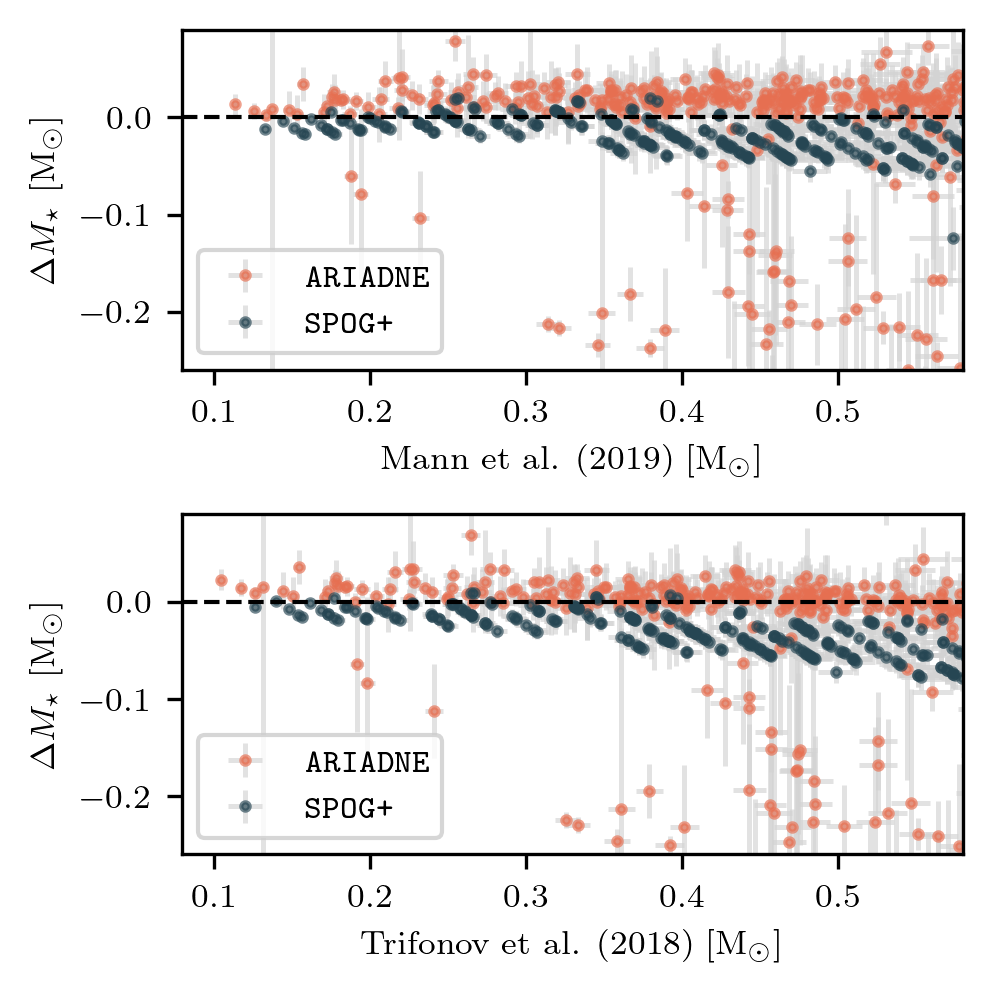}
\caption{Top panel: comparison between the masses calculated using the \texttt{M\_-M\_K-} code by \citet{Mann2019} and masses from \texttt{ARIADNE} and \texttt{SPOG+}. Bottom panel: comparison with the masses calculated using the polynomial mass--luminosity relation by \citet{Trifonov2018}. The ordinate of all panels shows the mass estimates by \texttt{ARIADNE} and \texttt{SPOG+} minus the mass--luminosity relation value for a given target.
\label{fig:general}}
\end{figure}

A comparison of these relations with the masses derived by \texttt{ARIADNE} and \texttt{SPOG+} is shown in \autoref{fig:general}. The \texttt{SPOG+} masses (shown as dark-blue points) cluster around the $0.025\,M_\odot$ grid spacing of the stellar models used by the code.

\texttt{ARIADNE} does not exhibit grid discretization effects but produces a group of significant outliers visible in both panels of \autoref{fig:general}. We find that for all outliers the tool has assigned ages $\lesssim 100$ Myr, and they correspond to the PMS population identified in the top panel of \autoref{fig:tic_mcomp}. Their overestimated masses likely arise from the age–mass degeneracy affecting isochrone fitting. These outliers are therefore excluded from the regression analysis. After removing these objects, we find that the \texttt{ARIADNE} masses agree with the \texttt{M\_-M\_K-} estimates with a slope consistent with unity but exhibit a systematic offset of $0.018 \pm 0.002\,M_\odot$. When compared with the \citet{Trifonov2018} relation, the regression yields a slope of $0.919 \pm 0.014$ and an offset of $0.037 \pm 0.007\,M_\odot$.

Although the pairwise correlations between catalogs are generally strong in the low-mass regime, similar systematic slopes and offsets are present when comparing \texttt{ARIADNE} and \texttt{SPOG+} with TICv8.2. For this reason, we include in our catalog additional mass estimates derived using the \texttt{M\_-M\_K-} code and the \citet{Trifonov2018} mass–luminosity relation for all low-mass stars with available $K_s$ magnitudes and distances from \citet{Bailer-Jones2021}.

\subsubsection{Massive Stars}

\begin{figure}[ht!]
\includegraphics[width=\columnwidth]{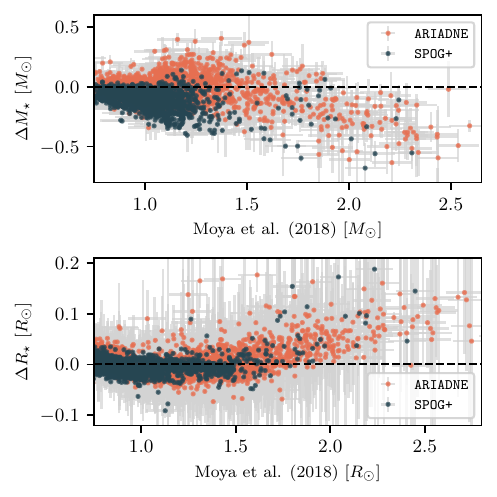}
\caption{Top panel: comparison between the masses by \texttt{ARIADNE}, \texttt{SPOG+} and those derived by the empirical relation by \citet{Moya2018}. The ordinate of both panels shows the values by \texttt{ARIADNE} and \texttt{SPOG+} minus the empirical relation value for a given target. Bottom panel: comparison between the radii estimations by \texttt{ARIADNE}, \texttt{SPOG+} and those derived using the relation by \citet{Moya2018}. 
\label{fig:moya}}
\end{figure}

To check the accuracy of our estimated parameters for the more massive stars, we used the empirical relations provided by \citet{Moya2018}. In this work, the authors analyze datasets from 11 different studies, providing stellar parameters derived from asteroseismology, eclipses in detached binaries, and interferometry. They build a sample of 934 stars, both main-sequence and evolved, of which  735 are spectral types F and G. This sample is then used to calibrate empirical relations for deriving stellar mass and radius. Since in our sample $T_\mathrm{eff}$, $L_\star$, and $[Fe/H]$ have the most reliable estimations from SED fitting and spectroscopy, we choose  the following relation:
\begin{equation}
    M_\star = a + b T_\mathrm{eff} + c L_\star + d [Fe/H],
\end{equation}
where the coefficients are \mbox{$a = (-3.16\pm0.10)\times 10^{-1}$}, \mbox{$b = (2.289\pm0.018)\times 10^{-4}$}, \mbox{$c = (3.88\pm0.05)\times 10^{-2}$}, and \mbox{$d = (1.31\pm0.04)\times10^{-1}$}. Since the majority of stars used to calibrate these relations are F and G stars, mass estimates above \mbox{$\sim1.6 M_\odot$} could be affected due to poor statistical sampling in this mass regime in the calibration sets. The results of this comparison can be seen on the top panel of \autoref{fig:moya}.

We see that there is an offset for all \texttt{SPOG+} masses of $-0.11\pm0.01\,M_\odot$, while \texttt{ARIADNE} is consistent with the empirical relations in the solar-like mass regime. Both of our methods differ from \citet{Moya2018} for very massive stars, likely due to the small number of stars in this mass range in the empirical calibration sample.

\subsection{Stellar Radii}

The high precision of Gaia DR3 parallaxes allows both \texttt{ARIADNE} and \texttt{SPOG+} to derive well-constrained stellar radii. In the lower-left panel of \autoref{fig:correls} we present the pairwise correlations for the sample of 278 stars with radius estimates available in all databases considered in this work. The radii are strongly correlated across all catalogs, with the lowest Pearson correlation coefficient being 0.9634 between \texttt{ARIADNE} and the \url{exoplanet.eu} database.

\citet{Moya2018} also provide an empirical relation for stellar radii in the following form:
\begin{equation}
    \log R_\star = a + b T_\mathrm{eff} + c \log{L_\star} + d [Fe/H],
\end{equation}
where the coefficients are \mbox{$a = (6.64\pm0.06)\times 10^{-1}$}, \mbox{$b = (-1.141\pm0.010)\times 10^{-4}$}, \mbox{$c = (4.617\pm0.013)\times 10^{-1}$}, and \mbox{$d = (9.2\pm1.4)\times10^{-3}$}. The comparison between this relation and the radii derived by \texttt{ARIADNE} and \texttt{SPOG+} can be seen on the bottom panel of \autoref{fig:moya}. Estimates from both methods diverge from the empirical relation for giants, possibly due to the low number of giant stars in the calibration sample of \citet{Moya2018}.

When comparing TICv8.2 with the radii derived from \texttt{ARIADNE} and \texttt{SPOG+} using the full sets of overlapping stars, we find no significant slopes or systematic offsets. Among the stellar parameters analyzed in this study, the stellar radius shows the highest level of agreement between the different catalogs.

\subsection{Stellar Surface Gravities}

\texttt{ARIADNE} provides a surface gravity estimate derived directly from its SED-fitting procedure. However, this quantity is sensitive to the adopted prior on $\log g$ and, when wide priors are used, the method tends to produce systematically underestimated values \citep{Vines&Jenkins2022}. For this reason, in addition to the SED-derived $\log g$, we also compute surface gravities using the stellar radius obtained from SED fitting and the stellar mass derived from isochrone interpolation within \texttt{ARIADNE}. We adopt this latter value as our primary $\log g$ estimate and use it in the comparisons with other catalogs.

The lower-right panel of \autoref{fig:correls} presents the pairwise correlations of $\log g$ for the 170 stars that have surface-gravity estimates available in all databases considered in this study. Among these sources, only \citetalias{Perdelwitz2024} provides direct spectroscopic measurements of $\log g$. Its values show systematically lower correlation coefficients with the other catalogs, including the parameters listed in the \url{exoplanet.eu} database. Even though these spectroscopic surface gravities are adopted as input priors for both \texttt{ARIADNE} and \texttt{SPOG+}, the resulting estimates from the two tools remain mutually consistent and agree well with the values reported by the remaining catalogs.

\section{Summary and Conclusions} \label{sec:highlight}

In this work, we present a new publicly available homogeneous catalog of stellar atmospheric and fundamental physical parameters for 5533 single stars monitored by major spectroscopic facilities. 
The sample combines targets from the {\sc HARPS-RVB}ank, the HIRES/Keck RV catalogs, and the {\sc CARMENES-DR1} sample, and it can be found in CDS(TBA), or in \mbox{\url{https://exo-restart.com/stellar-parameters-catalog}}. 
The catalog is intended to provide an internally consistent stellar reference sample for the analysis of RV exoplanet host stars, particularly for studies based on long-term Doppler surveys and, where applicable, combined RV-transit analyses.

We derive homogeneous stellar parameters using two independent stellar-evolution model grids. 
The first is the Bayesian SED-fitting framework \texttt{ARIADNE}, which combines multi-band photometry with stellar atmosphere models and MIST isochrones. The second is \texttt{SPOG+}, a Bayesian inference tool that derives stellar parameters by interpolating PARSEC evolutionary tracks. Input priors for these tools were constructed using spectroscopic parameters from \citetalias{Perdelwitz2024}, where available, and supplemented with values from TICv8.2 and the Gaia DR3 GSP-Aeneas pipeline. 
The resulting catalog, therefore, provides a consistent set of stellar parameters derived using two complementary methodologies, while retaining the ability to compare the behavior of the two models. 

We successfully obtained \texttt{ARIADNE} parameters for 5014 single stars and \texttt{SPOG+} parameters for 4018 single stars, covering a broad range of stellar types, including low-mass dwarfs, solar-type stars, sub-giants, and evolved stars. A comparison with Gaia DR3, TICv8.2, and the \url{exoplanet.eu} web-database shows good overall agreement. Effective temperatures are generally consistent, with minor systematic offsets for cool main-sequence and horizontal-branch stars. Surface gravities from \texttt{ARIADNE} (computed from isochrone-based masses and SED radii) and \texttt{SPOG+} agree with external catalogs, with only small offsets for high-gravity dwarfs. Stellar radii are tightly constrained and show excellent agreement across all catalogs, benefiting from precise Gaia DR3 parallaxes.

The largest method-dependent differences are found in stellar masses, as expected. For evolved stars, both \texttt{ARIADNE} and \texttt{SPOG+} tend to yield masses that are slightly higher than those listed in TICv8.2, while this offset is not clearly present in the comparison with the \url{exoplanet.eu} database. This likely arises because published values often rely on isochrone fitting using stellar models similar to \texttt{ARIADNE} and \texttt{SPOG+}. Both methods exhibit a slight offset for low-mass stars, and in some isolated cases \texttt{ARIADNE} misidentifies the evolutionary stages of stars, leading to inaccurate mass estimates. To mitigate these effects, we include empirical masses from \citet{Mann2019} and \citet{Trifonov2018} for low-mass stars. These empirical estimates are expected to be more robust for M dwarfs and other low-mass main-sequence stars.

Our parameter estimation approach with \texttt{ARIADNE} and \texttt{SPOG+} characterizes stellar metallicity through the global [Fe/H] abundance. Although this does not capture the full chemical abundance patterns that may be relevant for exoplanet-host studies, deriving homogeneous elemental abundances for the full sample would require dedicated spectroscopic analyses beyond the methodology adopted here.

In summary, this work presents one of the largest uniformly derived stellar-parameter catalogs for stars monitored by high-resolution, high-precision Doppler spectrographs. It also establishes the range of validity, applicability, and limitations of the two stellar-parameter homogenization tools employed across different stellar populations. The resulting catalog provides a robust basis for future studies of exoplanet occurrence rates, planetary bulk compositions, and the dynamical architectures of RV multiple-planet systems.

\begin{acknowledgments}
We thank the anonymous referee for constructive comments that improved the quality of this work. We are also grateful to Andreas Quirrenbach, Sabine Reffert, and Alex Golovin for helpful discussions and for their assistance during the preparation of this manuscript.

S.S., E.Z., D.R., D.S., D.A., V.B., and T.T. acknowledge support by the Bulgarian National Science Fund (BNSF) program ``VIHREN--2021'' project No. KP--06--DV/5/15.12.2021. L.T.-O. and J.T.T. acknowledge support from the Israel Science Foundation through grant 1404/22.
\end{acknowledgments}

\begin{contribution}

S.S. was responsible for writing and submitting the manuscript. E.Z. and
M.M. developed numerical tools for this project. D.R. contributed significantly 
to the analysis and tool development, and edited the manuscript. D.S., V.P., J.T.,
E.V.B., D.A., and V.B. provided essential research expertise and edited the manuscript. 
T.T. developed the initial research concept and edited the manuscript.

\end{contribution}

\facilities{ESO:3.6m(HARPS), Keck:I(HIRES), CAO:3.5m(CARMENES)}

\software{\texttt{astropy} \citep{2013A&A...558A..33A,2018AJ....156..123A,2022ApJ...935..167A}; \texttt{ARIADNE} \citep{Vines&Jenkins2022}; \texttt{SPOG+} \citep{Stock2018}; \texttt{SPECIES} \citep{speciesI_2018,speciesII_2021}; \texttt{dustmaps} \citep{Green2018}; \texttt{M\_-M\_K-} \citep{Mann2019}; \texttt{scipy} \citep{scipy_2020}; \texttt{TOPCAT} \citep{Taylor2011}.
}

\bibliography{Bibliography}{}
\bibliographystyle{aasjournalv7}

\end{document}